# Nuclear Spectroscopy for the Exploration of Mars and Beyond

*A White Paper submitted to the Planetary Science Decadal Survey 2023-2032*

**Authors:**

Katherine E. Mesick, Patrick J. Gasda
*Los Alamos National Laboratory, Los Alamos, NM 87545*

Travis S.J. Gabriel, Craig Hardgrove
*Arizona State University, Tempe, AZ 85287*

William C. Feldman
*Planetary Science Institute, Tuscon, AZ, 85719*

**Primary Author Contact:**

Katherine E. Mesick
Los Alamos National Laboratory
505-667-8273
kmesick@lanl.gov

**Signatories:**

Dan Coupland, Los Alamos National Laboratory
Suzanne Nowicki, Los Alamos National Laboratory
Oz Pathare, Planetary Science Institute

## Executive Summary

- **Nuclear spectroscopy will continue to play a critical role in planetary exploration in the next decade, including**
  - **Searching for habitable environments and biosignatures,**
  - **Identifying ideal targets for in-situ resource utilization, and**
  - **Continuing its track record in understanding planetary evolution.**
- **Nuclear spectroscopy is *the only technique* that provides bulk geochemical abundance at depth, at a range of spatial scales depending on platform.**

## Introduction

**Nuclear spectroscopy has played a critical role in planetary exploration for decades as it is the only instrumentation that provides bulk geochemical constraints at depth (up to one meter in the surface). Importantly, these instruments can identify and quantify water and other key elements relevant to assessing planetary volatile abundances and evolution, the search for extraterrestrial life, and in-situ resource utilization.** Planetary nuclear spectroscopy is the measurement of neutron and/or gamma-ray radiation signatures from planetary bodies, generated passively from either galactic cosmic ray (GCR) interactions within the surface or by active interrogation with a pulsed-neutron generator (PNG). GCR-induced ('passive') nuclear spectroscopy measurements can be obtained from any orbit within a few planetary radii of planetary bodies that have tenuous atmospheres, and on the surface. Active interrogation instruments (enabled by the presence of a PNG) can only be deployed on the surface and are particularly useful for planetary bodies with a substantial atmosphere. The non-invasive (truly *in-situ*) geochemical assays provided by complementary gamma-ray and neutron spectrometers (GRNS) are highly complementary to common photon-based spectroscopic/geochemical instrumentation.

Nuclear spectrometers provide geochemical maps at a variety of spatial scales. Previous GRNS instruments have largely flown on orbital missions, providing robust global maps of elemental abundances and quantification of H distributions. Orbital spectrometers provide global to regional geochemical maps at the ~10s-100s of km scale [*e.g.* Lawrence2010], such as the Mars Odyssey Neutron Spectrometer (MONS), which has an orbital altitude of 400 km and spatial resolution of 550 km [Maurice2011] and Lunar Prospector (LP), which had an orbital altitude as low as 30 km and spatial resolution of 45 km [Lawrence2004]. Orbital GRNS are particularly useful for compositional maps, identifying landing sites, and constraining global processes. Surface nuclear spectrometers have a much smaller footprint; the only such instrument deployed on Mars (the Dynamic Albedo of Neutrons (DAN) instrument) has a ~1 meter footprint, which allows for geochemical data at the local scale. The continued deployment on GRNS from orbit and in landed or rover configurations is critical to tie sub-regional compositional data to the global scale.

*To summarize, the unique nature of GRNS measurements for planetary remote sensing provide the following comparative advantages for planetary exploration:*

- **GRNS can identify and quantify water/H-rich deposits in the subsurface.**
- **GRNS provides *bulk* elemental assay of major and minor rock-forming elements.**
- **GRNS is sensitive to depths up to one meter in the subsurface.**
- **GRNS has a range of spatial sensitivity, from ~10-100 km in orbit to ~1 m on the surface.**
- **GRNS is a non-destructive technique: the rocks/regolith require no preparation and are not disturbed.**

*Nuclear spectrometers have a unique sensing depth, spatial coverage, and sensitivity to bulk composition and surficial or buried water/H deposits, making this technique highly complementary to common remote sensing methods.*

# Scientific Advancement Enabled by Nuclear Spectroscopy

In the past decade, significant advancements in our understanding of planetary processes have been enabled by nuclear spectroscopy. An abbreviated summary highlighting some of these scientific advancements is provided to illustrate the importance of GRNS measurements.

**Moon:** Global geochemical maps from LP provided definitive detection of three major, geologically distinct terranes and Th-rich provinces, providing new insight into lunar crustal differentiation and hot-spot volcanism [*e.g.* Lawrence1999, Jolliff2000]. Large enrichments of water-ice in permanently shadowed regions (PSRs) at the poles were detected with neutron spectrometers on LP and LRO [*e.g.* Feldman1998, Lawrence2006, Sanin2017].

**Mars:** MONS provided the first detailed, global maps of the vertical distribution of hydrogen in the top meter of the surface, including detection and characterization of shallowly-buried water-ice at mid-high latitudes, consistent with recent ice-exposing craters [*e.g.* Pathare2018]. On the surface in Gale crater, the DAN instrument has advanced our understanding of magmatic processes through the characterization of a silicic volcaniclastic layer [*e.g.* Czarnecki2020].

**Mercury:** The MESSENGER GRNS confirmed the presence of water-ice in PSRs at the north pole, and further, characterized the abundance and layering of water ice and total mass to be consistent with delivery from comets or volatile-rich asteroids [*e.g.* Lawrence2013]. In concert with other instruments, the MESSENGER GRNS inferred the presence of a carbon-rich layer indicative of crust formed early in Mercury's evolution atop a magma ocean [*e.g.* Peplowski2015].

**Asteroids:** Elemental abundances measured by the DAWN GRNS confirmed that Vesta is likely the parent body of HED meteorites, however, H-enrichments were observed indicating an exogenic source of H from the infall of carbonaceous chondrite material [*e.g.* Prettyman2012]. Global mapping by the same instrument confirmed Ceres as an icy body [*e.g.* Prettyman2017].

> *A considerable number of high-profile advancements in planetary science have been enabled by nuclear spectroscopy instruments.*

# Nuclear Spectroscopy for Exploration in the Next Decade

Important planetary science and human exploration goals for rocky and icy bodies in our solar system can be addressed through the deployment of nuclear spectrometers on future missions. These goals include critical science questions related to planetary evolution, the search for life, and the detection of volatiles. Thus, the combination of currently planned missions and the deployment of new nuclear spectrometers will continue to enable high-profile scientific advancement in the coming decades. We cannot comprehensively discuss all of the important science questions and potential missions that GRNS would be beneficial to in the next decade within the scope of this white paper, thus we provide focused examples for two topics.

## *Nuclear Spectroscopy and the Search for Life*

In the search for evidence of past or present life, **nuclear spectroscopy techniques provide key bulk geochemical context necessary for the characterization of potentially habitable modern and ancient environments, and biosignatures, if found**. GRNS can aid in answering several high-priority science objectives related to this topic, such as Goal I described in MEPAG's Mars Science Goals, Objectives, Investigations, and Priorities 2020 document [MEPAG2020]: *"Determine if Mars ever supported, or still supports, life."* Importantly, neutron measurements can detect the location, extent, and quantity of H (as hydrated minerals or ice) within a meter of the surface, and gamma-ray measurements can detect and quantify important elements such as bioessential elements (C, H, N, O, P, S) and micro-nutrients (*e.g.* Fe, Mn, Ni, Zn). Due to this unique sensitivity, nuclear spectrometers are specifically called for in investigations to address high-priority sub-objectives within Goal I of MEPAG's 2020 Goals document. The utility of GRNS investigations extends beyond Mars to Icy Bodies and can address a range of astrobiological questions as outlined in Question 5 of the NASA Astrobiology Strategy 2015 [Hayes2015]: *"Identifying, Exploring, and Characterizing Environments for Habitability and Biosignatures."*

In addition to the detection of water and bioessential elements, GRNS instruments can search for elements associated with certain geological environments where biological signatures are most likely to be found, such as hydrothermal systems, saline lakes, and caves. For landed or roving measurements, GRNS analysis can quickly identify targets of interest (*i.e.* identify bulk rocks with unique geochemical signatures) for follow-up investigation and is complementary to instruments that operate at smaller spatial and depth scales, such as x-ray spectrometers, Laser Induced Breakdown Spectroscopy (LIBS), or Raman spectroscopy. GRNS is also particularly useful for understanding whether smaller scale sample geochemical analyses are indicative of the bulk rock, which can inform selection of samples for analysis and sample return.

Some specific examples of the unique role GRNS would play in missions searching for biosignatures within the next decade are:

- **Identification and characterization of hydrothermal environments**, where in such settings on Earth, life can thrive and many prebiotic reactions likely took place at the dawn of life. For example, Ni is especially important to the origin of life at hydrothermal vents on Earth [Martin2008, Sojo2016], and enhanced concentrations of Fe coinciding with Zn, Ni, Mn, and Cu could indicate a metal sulfide deposit indicative of a hydrothermal environment. Other hydrothermal settings such as hot springs or geysers often produce sinter deposits that are rich in amorphous silica (Si and H) and depleted in Fe, and, because of the proclivity for these rocks to preserve microbiological texture, they were identified as critical targets for the Mars exploration program [Farmer1999]. Potential hot-spring deposits were identified at the Spirit landing site Home Plate [Ruff2020] and have been identified from orbit at the Mars 2020 Perseverance rover landing site in Jezero crater [Tarnas2019]. GRNS instruments have proven to be highly sensitive to the geochemical indicators of hydrothermal deposits and their distribution in the subsurface, as was demonstrated during exploration of several subsurface high-silica deposits and a hematite-bearing ridge with the active neutron spectrometer instrument DAN, onboard the Curiosity rover in Gale crater [Rapin2020, Czarnecki2020]. Since the follow-up exploration and characterization of silica sinter deposits at the Spirit, Curiosity, and Perseverance rover landing sites may be important in the coming decade, GRNS instruments would play a key role in their identification, characterization, and mapping, especially at depth.
- **Identification and mapping of evaporite deposits**, another key location to look for preserved biological materials since they typically form in a lacustrine or marine setting, thus are locations that likely had abundant water in the past before drying. On Earth, extremophiles adapted to live in the highly saline and highly acidic or alkaline conditions of drying lakes, and even within the evaporite deposits themselves [*e.g.* Wierzchos2006]. These deposits have a high potential to preserve organics or can harbor extant life, and are associated with the presence of elements such as Cl, S, and B, all of which have been detected on Mars [*e.g.* Gasda2017]. GRNS is extremely sensitive to these elements in the bulk and at depth, thus the deployment of such instruments is ideal for characterization and mapping of these deposits.
- **Mapping and characterization for subsurface exploration**, such as lava tubes or caves. For example, given that the surface of Mars is poorly habitable in the present day, the martian subsurface has been proposed as a high-value target for future exploration and the search for life [Northup2011, Carrier2020]. The subsurface of Mars is protected from the harsh radiation environment that exists due to the tenuous present-day atmosphere. Furthermore, the subsurface is closer to internal heat sources and is closer to potential reservoirs of ice and liquid brines, with stability depending on the latitude and depth. Deep drilling missions aim to tap into deep water aquifers on Mars in search of extant life, with the necessary drilling technology development proposed in the next decade [*e.g.* Stamenković2019]. Subsurface caves offer easier and direct access to the shallow subsurface and have been proposed as a potential access point to explore habitable settings [*e.g.* Léveillé2010, Carrier2020]. In a similar fashion to surface exploration, a GRNS instrument would be ideal to map the bulk elemental composition of the interior of subsurface lava tubes or caves. For example, a neutron spectrometer would detect pockets of water or ice within the wall of a cave that warrant additional observation by other instruments on a mission. A gamma-ray spectrometer would be used to produce bulk chemical maps of the cave interior for geological context of habitability and to help identify astrobiological targets of interest for follow-up sample collection.

*Preparing for Human Exploration*

In-situ resource utilization (ISRU) will play an important role in the next decade, as humans prepare for in-situ exploration of the Moon and Mars. **Nuclear spectroscopy is the only technique that can quantify the abundance of critical resources at depths that are easily accessible by human missions** (*i.e.* in the top meter of the surface). The 2018 NASA Strategic Plan [NASASP2018] Strategic Objective 2.2 outlines goals for human exploration into deep space, including to the surface of the Moon and eventually Mars. For the Moon, the Lunar Exploration Roadmap (LER): Exploring the Moon in the 21$^{st}$ Century: Themes, Goals, Objectives, Investigations, and Priorities, 2016 document [LER2016] describes multiple Objectives related to ISRU that nuclear spectrometers can address. This includes mapping and characterization of polar cold traps and characterization of other materials relevant to ISRU which may be present in crater ejecta. The importance of searching for accessible water is also echoed for Mars exploration in MEPAG's 2020 Goals document [MEPAG2020] Goal 4: *"Prepare for human exploration"* and MEPAG's Next Orbiter Science Analysis Group (NEX-SAG) report [NEXSAG2015] strategic knowledge gaps RS-A: *"Find and quantify the extent of shallow ground ice within a few meters of the surface and characterize its ice-free overburden"* and RS-B: *"Identify deposits with hydrated minerals as a water resource, and potential contaminants within these deposits."* Specifically, MEPAG Goal 4 Objective C calls for investigations characterizing potentially extractable water resources, either at low latitudes where water may be bound to minerals or at high latitudes where it likely exists as water ice, similar to the NEXT-SAG goals. While neutron spectrometers alone can determine the abundance of hydrogen, simultaneously deploying a gamma-ray spectrometer provides additional information on the form the hydrogen may be in (*e.g.* bound to minerals or as ice) and other contaminants that may be present. Gamma-ray spectrometers also provide information on elemental abundance of other materials relevant to ISRU, such as fuel sources (*e.g.* Th, U) and possible building materials (*e.g.* Al, Ti, Fe, Mg, Si).

Several upcoming missions include the mapping of H resources on the Moon, however, additional future payloads should consider GRNS instruments to bridge important gaps in spatial scale in the human exploration setting. Mapping of hydrogen in cold traps on the Moon are science objectives of upcoming missions that include neutron spectrometers: LunaH-Map [Hardgrove2020], which will map hydrogen from low-altitude passes over the South pole, and Viper [Ennico-Smith2020], a rover which will explore the South pole. Importantly, these missions extend the maps of H from previous orbital missions to specific areas on the Moon. Beyond these, some examples of missions where GRNS would be beneficial in the next decade are:

- **Assessing rock/ice composition with penetration or drilling experiments.** Such a mission could deploy passive (top few meters) or active nuclear spectrometers within the penetrator body [*e.g.* PaigeWP] or down a drilled borehole to assess subsurface hydrogen deposits and elemental composition with depth. GRNS instruments have served as a principal tool for borehole characterization in oil and gas exploration on Earth [*e.g.* Hertzog1979]. GRNS would be useful for characterizing in a detailed manner resources available in cold traps on the Moon or within layered ice deposits on Mars.
- **Producing high-spatial resolution maps of resources on Mars.** On Mars, higher spatial resolution regional to global maps of potential water deposits, than those obtained by traditional nuclear spectrometers in orbit, are required to make important advances in the next decade for informing ISRU landing site selection. Obtaining such measurements can be achieved by: 1) advancing instrumentation to allow higher resolution mapping from standard orbital platforms, 2) deploying nuclear spectrometers on novel, low-altitude platforms such as a balloon or a low-altitude orbiter (providing kilometer to tens of kilometer scale spatial resolution), or 3) deploying additional nuclear spectrometers on landers and rovers in regions of interest (providing meter scale spatial resolution).

## Summary

In this white paper, we have highlighted examples of significant scientific advancement enabled by nuclear spectroscopy, and argued for the continued deployment of nuclear spectroscopy instruments in the next decade due to the unique information this technique provides. **GRNS is the only technique that describes bulk geochemistry within the top meter of the surface, in a truly in-situ (non-destructive) manner, with spatial scales that can bridge the gap between local, regional, and global processes**. Gamma-ray and neutron spectrometers are themselves highly complementary and should be deployed together, and **GRNS are highly complementary to other remote sensing methods that operate at different depth and/or spatial scales**. As such, nuclear spectrometers should continue to be deployed on surface and orbital platforms, to a variety of target bodies. On landed platforms in particular, GRNS are particularly useful for understanding whether smaller scale sample geochemical analyses performed by other instruments are indicative of the bulk rock and can inform sample return selection.

We provided two specific examples of important science themes that GRNS can contribute to in the next decade: **characterization of potentially habitable modern and ancient environments, and biosignatures,** and **quantifying the abundance of critical resources at depths that are easily accessible by human missions**. However, as illustrated by the scientific advancements enabled by previous GRNS instruments, **the utility of nuclear spectroscopy has a much broader impact in understanding planetary processes, and should not be overlooked**.

Some targeted instrument development for GRNS in the next decade that will enable this technique to provide improved information are: 1) Development of lower resource (size, weight, power) GRNS to make this technique accessible to a wider range of missions, such as small sats, micro rovers, or low-altitude platforms, and/or 2) Development of GRNS instruments able to achieve higher spatial resolution from orbital platforms.